\documentclass[aps,twocolumn,showpacs,preprintnumbers]{revtex4-1}

\usepackage{amsmath,amssymb,bm}
\usepackage[dvips]{graphicx}
\usepackage{yfonts}
\newcommand{\beq}{\begin{eqnarray}}
\newcommand{\eeq}{\end{eqnarray}}
\renewcommand\d{\partial}
\begin{document}

\preprint{YITP-13-9}
\title{Chiral Plasma Instabilities}
\author{Yukinao Akamatsu}
\affiliation{Kobayashi-Maskawa Institute for the Origin of Particles and the Universe,
Nagoya University, Nagoya 464-8602, Japan}
\author{Naoki Yamamoto}
\affiliation{Yukawa Institute for Theoretical Physics,
Kyoto University, Kyoto 606-8502, Japan}
\affiliation{Maryland Center for Fundamental Physics, 
Department of Physics, University of Maryland,
College Park, Maryland 20742-4111, USA}
\begin{abstract}
We study the collective modes in relativistic electromagnetic or quark-gluon 
plasmas with an asymmetry between left- and right-handed chiral fermions, 
based on the recently formulated kinetic theory with Berry curvature corrections.
We find that there exists an unstable mode, signaling the presence of a plasma 
instability. We argue the fate of this ``chiral plasma instability"  including the 
effect of collisions, and briefly discuss its relevance in heavy ion collisions and 
compact stars.
\end{abstract}
\maketitle

\emph{Introduction.}---%
Parity violating effects related to quantum anomalies play an important 
role in a wide range of physics from quantum Hall systems to cosmology. 
One example in the transport phenomena is the parity violating current in the 
presence of a magnetic field and an asymmetry between left and right-handed 
fermions, parametrized by the chiral chemical potential $\mu_5 \equiv \mu_R - \mu_L$. 
This is called the chiral magnetic effect (CME)
\cite{Vilenkin:1980fu, Nielsen:1983rb, Alekseev:1998ds, Son:2004tq, Fukushima:2008xe}.
Recently hydrodynamics \cite{Son:2009tf} and kinetic theory
\cite{Son:2012wh, Zahed:2012yu, Stephanov:2012ki, Son:2012zy, Chen:2012ca} 
have been appropriately modified to describe quantum anomalies and the CME 
(see also Refs.~\cite{Loganayagam:2012pz, Gao:2012ix});
in the kinetic theory, essential corrections are the Berry curvature, 
the concept diversely applied in condensed matter physics \cite{Xiao:2010, Volovik}.
These developments enable us to systematically understand the properties and 
dynamical evolution of various systems with hitherto neglected anomalous effects.

In this Letter, we study the collective modes and their consequences in relativistic 
electromagnetic or quark-gluon plasmas (QGP) at finite $\mu_5$ \cite{mu_5}
and temperature $T$ based on the new kinetic theory.
(For a review on the collective modes \emph{without} Berry curvature 
corrections, see Ref.~\cite{Blaizot:2001nr}.)
We show that the Berry curvature corrections dramatically modify the 
dispersion relation of the collective modes, and in particular, lead to unstable modes, 
signaling the presence of a plasma instability. We shall call it the chiral plasma instability. 
Closely related instability was studied within the electroweak theory 
at large lepton chemical potential \cite{Redlich:1984md, Rubakov:1985nk} 
and in the context of the early Universe at $T \gg \mu_5$ 
\cite{Joyce:1997uy, Laine:2005bt, Boyarsky:2011uy}
where it is ascribed to a possible origin of the primordial magnetic field.

Our main purpose of this Letter is to reveal the potential importance 
of the chiral plasma instabilities in heavy ion collisions and compact stars.
It has been argued that the QGP created in noncentral heavy ion collisions 
may contain locally finite $\mu_5$ for quarks and a large external 
magnetic field to yield an observable CME 
\cite{Kharzeev:2007tn, Kharzeev:2007jp, Fukushima:2008xe}. 
It was also suggested that a degenerate electromagnetic plasma with finite 
$\mu_5$ for electrons may exist inside neutron stars due to the 
parity violating weak process, where the CME is generated in a strong 
magnetic field \cite{Charbonneau:2009ax}. 
We show within the kinetic theory that these QCD and QED plasmas
at finite $\mu_5$ are dynamically unstable and reduce $\mu_5$ by 
converting it to (color) electromagnetic fields with magnetic helicity.
We estimate the typical time scales of the chiral plasma instabilities 
for the QCD and QED plasmas as
$\tau_{\rm QCD} \sim 1/(\alpha_s^2 \mu_5 \ln \alpha_s^{-1})$ 
for $\mu_5\sim T$ and $\tau_{\rm QED} \sim 1/(\alpha^2 \mu_5$),
respectively.

\emph{Kinetic theory with parity violating effects.}---%
Let us describe a chiral plasma within the kinetic framework.
We first consider the collisionless kinetic theory and we shall argue the 
effect of collisions later.
We also consider the regime of a sufficiently weak gauge field $A^{\mu}$,
where there is no essential difference between Abelian and non-Abelian 
gauge fields, up to color and flavor degrees of freedom 
\cite{Romatschke:2003ms, Arnold:2003rq}. 
For simplicity of notation, we first consider QED, and we will 
generalize it to QCD later (with some modifications mentioned below).
We assume massless quarks and spatially homogeneous $\mu_5$ below.

Recall the Maxwell equation
\beq
\label{eq:maxwell}
\d_{\nu} F^{\nu \mu} = j_{\rm ind}^{\mu} + j_{\rm ext}^{\mu},
\eeq
where $j_{\rm ind}^{\mu}$ is the induced current 
and $j_{\rm ext}^{\mu}$ is the external current.
For the small gauge  field $A^{\nu}$, 
the induced current can be expressed, via the linear response theory, as
\beq
\label{eq:LRT}
j_{\rm ind}^{\mu}(K) = \Pi^{\mu \nu}(K) A_{\nu}(K),
\eeq
in the momentum space, where $\Pi^{\mu \nu}$ is the (retarded) 
self-energy and $K^{\mu}=(\omega, {\bf k})$ is the four-momentum.
From Eqs.~(\ref{eq:maxwell}) and (\ref{eq:LRT}), we obtain
\beq
\label{eq:dispersion} 
[K^2 g^{\mu \nu} - K^{\mu} K^{\nu} + \Pi^{\mu \nu}(K)]A_{\nu}(K) = - j_{\rm ext}^{\mu}.
\eeq
Because of the gauge invariance, which allows us to shift 
$A_{\nu} \rightarrow A_{\nu} + K_{\nu}$,
we can choose the temporal gauge $A_0=0$. Then
Eq.~(\ref{eq:dispersion}) can be rewritten, using the electric field, as
\beq
\label{eq:eigen}
[\Delta^{-1}]^{ij}E^j \equiv 
[(k^2 - \omega^2)\delta^{ij} - k^{i} k^{j} + \Pi^{ij}]E^j = 
- i \omega  j_{\rm ext}^{i},
\nonumber \\
\eeq
where $k \equiv |{\bf k}|$.
The dispersion relation for the collective modes in the system
can be found by computing the poles of $\Delta^{ij}$; 
a mode with the dispersion relation $\omega=\omega(k)$ 
satisfying ${\rm Im}(\omega)>0$, if exists, implies an instability.

The explicit form of $\Pi^{\mu \nu}$ including the parity violating effects 
can be found by using the kinetic theory with Berry curvature corrections 
\cite{Son:2012wh, Stephanov:2012ki, Son:2012zy}.
To illustrate the point and for simplicity, we take the initial equilibrium distribution 
function $n_{\bf p}^0$ to be isotropic (up to the Zeeman effect in the presence
of a magnetic field; see below). 
It should be remarked that though the same result for $\Pi^{\mu \nu}$ 
may be obtained in perturbation theory for the isotropic $n_{\bf p}^0$ \cite{Son:2012zy} 
(see also Ref.~\cite{Kharzeev:2009pj}), the kinetic theory 
would be more versatile for this and later purposes;
the topological origin of the parity violating tensor $\Pi^{\mu \nu}_-$ will be apparent
in this framework. It is also easy to include the effect of collisions, as we shall do later. 
Moreover, it is suitable for other future applications, such as the inclusion of 
anisotropy of $n_{\bf p}^0$ and numerical simulations of the dynamical evolution 
of plasmas.

At the leading order in $A^{\mu}$, the kinetic theory is given by
\cite{Son:2012zy}
\beq
\label{eq:simple_kin}
\left({\d_t} + {\bf v} \cdot {{\bm \nabla}_{\! \bf x}} \right) n_{\bf p}
+\left(e{\bf E} + {\bf v} \times e{\bf B} - 
{{\bm \nabla}_{\! \bf x} \epsilon_{\bf p}} \right) \cdot 
{\bm \nabla}_{\! \bf p} n_{\bf p}=0,
\nonumber \\
\eeq
where ${\bf v}={\bf p}/p$,
$\epsilon_{\bf p}=p(1 - e{\bf B} \cdot {\bm \Omega}_{\bf p})$
with ${\bm \Omega}_{\bf p} = \pm {\bf p}/{2p^3}$ the Berry curvature 
for right- and left-handed fermions, respectively.
The energy of the chiral fermion is shifted from $\epsilon_{\bf p}=p$ 
by the amount $- p e{\bf B} \cdot {\bm \Omega}_{\bf p}$ due to 
the magnetic moment of chiral fermions at finite $\mu$ \cite{Son:2012zy}.
This correction makes the term involving ${\bm \nabla}_{\! \bf x} \epsilon_{\bf p}$
in Eq.~(\ref{eq:simple_kin}) nonvanishing for an inhomogeneous magnetic field,
unlike the conventional Vlasov equation. 
In the presence of the Berry curvature flux, the definition of the 
current is also modified to \cite{Son:2012wh, Stephanov:2012ki, Son:2012zy}
\beq
\label{eq:j}
{\bf j} = -e^2 \int_{\bf p}
\left[\epsilon_{\bf p}{\bm \nabla}_{\! \bf p} n_{\bf p}
+ \left({\bm \Omega}_{\bf p} \cdot {\bm \nabla}_{\!\bf p} n_{\bf p} \right) \epsilon_{\bf p} {\bf B}
+ \epsilon_{\bf p} {\bm \Omega}_{\bf p} \times {\bm \nabla}_{\! \bf x} n_{\bf p} \right].
\nonumber \\
\eeq

For a moment, we concentrate on right-handed fermions with chemical potential $\mu$. 
(This theory coupled to dynamical gauge fields itself is not well defined because of 
the gauge anomaly, but we shall eventually consider the theory with 
both right- and left-handed fermions, so that the gauge anomalies 
are cancelled out.)
We solve the linearized equation of (\ref{eq:simple_kin}) in terms of the deviation 
$\delta n_{\bf p}$ from the thermal equilibrium state 
$n_{\bf p}^0 = 1/[e^{(\epsilon_{\bf p}-\mu)/T} + 1]$,
where $n_{\bf p} = n_{\bf p}^0 + \delta n_{\bf p}$. 
Note that $n_{\bf p}^0$ does not obey $n_{\bf p}^0=n_{-{\bf p}}^0$ 
and breaks parity owing to the Zeeman effect.

Substituting the solution $n_{\bf p}$ to the current ${\bf j}$ and 
using the linear response theory (\ref{eq:LRT}), the self-energy of the gauge field 
can be expressed as $\Pi^{\mu \nu}(K) = \Pi^{\mu \nu}_+(K) +  \Pi^{\mu \nu}_-(K)$,
where \cite{Son:2012zy}
\begin{align}
\label{eq:P-even}
\Pi^{\mu \nu}_+(K) &= -m_D^2 \left[\delta^{\mu 0} \delta^{\nu 0} -
\omega \int \frac{d{\bf v}}{4\pi} \frac{v^{\mu} v^{\nu}}{v \cdot K + i \epsilon}\right], 
\\
\label{eq:P-odd}
\Pi^{i j}_-(K) &= \frac{e^2 \mu}{4\pi^2} i \epsilon^{ijk}k^k 
\left(1 - \frac{\omega^2}{k^2} \right)[1 - \omega L(K)],
\end{align}
are the parity-even and parity-odd self energies with
\begin{gather}
\label{eq:Debye}
m_D^2 = e^2 \left(\frac{T^2}{6} + \frac{\mu^2}{2\pi^2} \right), \\
\label{eq:Lindhard}
L(K) = \frac{1}{2k}\ln \frac{\omega + k}{\omega-k}.
\end{gather}
Here $i,j,k$ denote the spatial indices
[$\Pi^{\mu \nu}_-(K)$ is vanishing otherwise] and $v^{\mu}=(1, {\bf v})$.
Note that $\Pi_-$ is shown to be $T$ independent from the topological 
nature of the Berry curvature 
(see the similar computation of the CME in Ref.~\cite{Son:2012wh}).

\emph{Collective modes.}---%
We perform a tensor decomposition for the self-energy $\Pi^{ij}$.
Recall that $\Pi^{ij}$ is not symmetric with respect to the indices $i$ and $j$
due to the parity violating term $\Pi_-^{ij}$; 
we need to use not only the longitudinal and transverse projectors,
$P_L^{ij} = {\hat k}^i {\hat k}^j$ and $P_T^{ij} = \delta^{ij} - {\hat k}^i {\hat k}^j$,
but also the antisymmetric tensor, $P_A^{ij} = i \epsilon^{ijk} \hat k^k$,
to fully decompose the tensor structure of $\Pi^{ij}$, where $\hat k^i = k^i/k$.
Then $[\Delta^{-1}]^{ij}$ can be written as
\begin{gather}
\label{eq:inv}
[\Delta^{-1}]^{ij} = C_L P_L^{ij} + C_T P_T^{ij} + C_A P_A^{ij},\\
C_L= - \omega^2 + \Pi_L, \quad 
C_T= -\omega^2 + k^2 + \Pi_T, \quad
C_A= \Pi_A,  \nonumber
\end{gather}
where
\begin{subequations}
\begin{align}
\label{eq:Pi_L}
\Pi_L &= - m_D^2 \frac{\omega^2}{k^2}[1 - \omega L(K)], \\
\label{eq:Pi_T}
\Pi_T &= m_D^2 \frac{\omega^2}{2k^2} 
\left[ 1 + \frac{k^2 - \omega^2}{\omega} L(K) \right], \\
\label{eq:Pi_A}
\Pi_A &= \frac{e^2 \mu k}{4\pi^2} 
\left(1 - \frac{\omega^2}{k^2} \right)[1 - \omega L(K)].
\end{align}
\end{subequations}

In order to compute $\Delta^{ij}$ [the inverse of Eq.~(\ref{eq:inv})],
it is convenient to use the properties of the (mutually commutable) 
projectors,
\begin{subequations}
\begin{gather}
P_L^2 = P_L, \quad
P_T^2 = P_T, \quad
P_L P_T = 0, \\ 
P_A^2 = P_T, \quad
P_L P_A =0, \quad
P_T P_A = P_A, 
\label{eq:P_A}
\end{gather}
\end{subequations}
from which we obtain
\beq
\label{eq:Delta}
\Delta^{ij} = \frac{1}{C_L}P_L^{ij} + \frac{C_T}{C_T^2 - C_A^2} P_T^{ij} 
- \frac{C_A}{C_T^2 - C_A^2} P_A^{ij}.
\eeq
Therefore, the dispersion relations of collective modes are given by
$C_L=0$ and $C_T \pm C_A =0$, which, respectively, reduce to
\begin{align}
\label{eq:dispersion_L}
\omega^2 &= \Pi_L(K), \\
\label{eq:dispersion_T}
\omega^2 &= k^2 + \Pi_T(K) \pm \Pi_A (K).
\end{align}
While the dispersion relation for the longitudinal component is 
identical to the case without $\Pi_A$, that for the transverse component
is split up into two parts by the effect of $\Pi_A$.

Let us compute the dispersion relations of the collective modes
based on Eqs.~(\ref{eq:dispersion_L}) and (\ref{eq:dispersion_T}).

\begin{figure}[b]
\begin{center}
\includegraphics[width=5cm, angle=270]{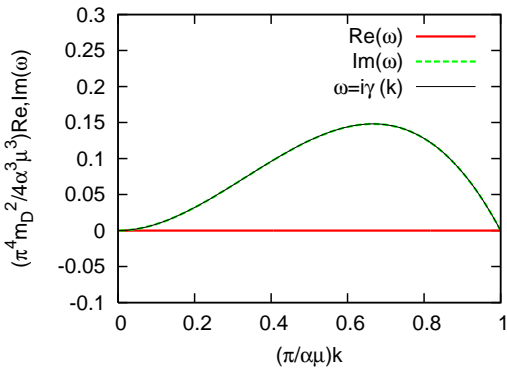}
\end{center}
\vspace{-0.5cm}
\caption{Real and imaginary parts of the dispersion relation 
$\omega=\omega(k)$ for the unstable mode.}
\label{fig:instability}
\end{figure}

In the long-wavelength limit $|\omega| \gg k$, 
$\Pi_A$ is negligible compared with $\Pi_{L,T}$ in the leading order, and 
\beq
\Pi_{L,T} = \frac{1}{3} m_D^2 \equiv \omega_{\rm pl}^2.
\eeq
The dispersion relations for the longitudinal and transverse modes are 
both given by $\omega^2 = \omega_{\rm pl}^2$,
implying that nonstatic gauge fields oscillate with the plasma frequency
$\omega_{\rm pl}$.

In the quasistatic limit $|\omega| \ll k$, using
$\omega L(K) =  \mp \frac{i \pi}{2} x
+ x^2 + \frac{1}{3} x^4 + \cdots$
for positive and negative Im$(\omega)$ with 
$x \equiv |\omega|/k \ll 1$,
the transverse dispersion relation can be obtained perturbatively in $x$. 
Equation~\eqref{eq:dispersion_T}  with the minus sign has the solutions,
\beq
\label{eq:unstable}
\omega = \pm i \gamma(k),
\quad 
\gamma(k) = \frac{4 \alpha \mu}{\pi^2 m_D^2} k^2\left(1-\frac{\pi k}{\alpha \mu}\right)
\eeq
for $0\leq k\leq \alpha \mu/\pi$ with $\alpha = e^2/(4\pi)$,
while that with the plus sign does not;
see Fig.~\ref{fig:instability} for the numerical result of Eq.~\eqref{eq:dispersion_T}
in this regime.
The mode $\omega= i \gamma(k)$ with $\gamma>0$ is unstable and grows 
exponentially as $e^{-i \omega t} \sim e^{\gamma t}$, 
indicating the plasma instability. 
For $\mu \sim T$, the typical scales of the unstable mode are 
\beq
\label{eq:scale_inst}
k \sim \alpha \mu, \qquad 
|\omega| \sim \frac{\alpha\mu k^2}{m_D^2} \sim \alpha^2 \mu,
\eeq
and the time scale of the plasma instability is
$\tau_{\rm inst} = 1/{\gamma}  \sim 1/(\alpha^2 \mu)$.

Without $\Pi_A$ originating from Berry curvature corrections, there was no unstable 
mode for isotropic $n_{\bf p}^0$. Note also that the criteria in Ref.~\cite{Arnold:2003rq}
for the conventional (Weibel) plasma instability \cite{Weibel:1959zz} is not applicable here. 
This is because the assumptions there, $n_{\bf p}^0 = n_{-{\bf p}}^0$ and $\Pi^{ij}=\Pi^{ji}$, 
are violated by the parity violating effects.

Now these results can easily be extended to a plasma 
at finite vector and chiral chemical potentials, 
$\mu_V = \mu_R + \mu_L$ and $\mu_5 = \mu_R - \mu_L$,
by the following replacements: 
$\mu^2 \rightarrow \mu_R^2 + \mu_L^2 = \frac{1}{2}(\mu_V^2 + \mu_5^2)$
in $m_D^2$, and $\mu \rightarrow \mu_R - \mu_L = \mu_5$ in $\Pi_-$, 
$\Pi_A$ and Eq.~(\ref{eq:unstable}).

\emph{The fate of chiral plasma instabilities.}---%
What is the fate of chiral plasma instabilities above?
Below we shall argue that this instability reduces $\mu_5$
so that it is weakened. 
We first recall the relation of triangle anomalies,
\beq
\label{eq:anomaly}
\d_t Q_5 = \frac{2 \alpha}{\pi} \int_{\bf x} {\bf E} \cdot {\bf B},
\eeq
where $Q_5$ is the global chiral charge. 
Equivalently,
\beq
\label{eq:helicity}
\d_t \left(Q_5 + \frac{\alpha}{\pi} {\cal H}\right)=0, \quad
{\cal H} = \int_{\bf x} {\bf A} \cdot {\bf B},
\eeq
where ${\cal H}$ is the Chern-Simons number 
(which is also called the magnetic helicity in plasma physics). 
In QCD, there is an additional $A^3$ term in ${\cal H}$, but it 
is negligible under the assumption of small $A$. 
Equation (\ref{eq:helicity}) is the conservation of the helicity.

We now show the reduction of $\mu_5$ within the kinetic theory
in the regime under consideration.
We concentrate on the unstable mode, $C_T \simeq C_A$ at $|\omega| \ll k$, 
for which the longitudinal part is negligible. Using Eq.~(\ref{eq:Delta}),
$A^i = \Delta^{ij} j^{j}_{\rm ext}$ [see Eq.~(\ref{eq:eigen})]
reduces to 
\beq
A^i(K) \simeq \frac{i \beta(k)}{\omega - i \gamma(k)} J_{\rm ext}^i(K),
\eeq
where we defined $\beta(k)=-2k/(\pi m_D^2)$ and
$J_{\rm ext}^i(K) \equiv (P_T^{ij} - P_A^{ij}) j^j_{\rm ext}(K)$.
In the $(t,{\bf k})$ space, $A^i$ satisfies 
\beq
[\d_t - \gamma(k)] A^i(t,{\bf k}) = \beta(k) J_{\rm ext}^i(t,{\bf k}).
\eeq
This can be solved in terms of $A^i$ for 
$j_{\rm ext}(t,{\bf k})=j_{\rm ext}({\bf k})\delta(t)$:
\beq
A^i(t,{\bf k}) = \beta e^{\gamma t}\theta(t) J^i_{\rm ext}({\bf k}).
\eeq
We then obtain $E^i = - \d_t A^i $ and $B^i = -k P_A^{ij} A^j$ as
\begin{align}
E^i(t,{\bf k}) & = -\beta [\gamma e^{\gamma t} \theta(t) + \delta(t)] J^i_{\rm ext}({\bf k}),
\\ 
B^i(t,{\bf k}) & = k \beta e^{\gamma t} \theta(t) J^i_{\rm ext}({\bf k}),
\end{align}
where we used $P^A J_{\rm ext} = -J_{\rm ext}$ that follows from
Eq.~(\ref{eq:P_A}).
Hence we find ${\bf E}(t,{\bf k}) \cdot {\bf B}^*(t,{\bf k})<0$
and ${\bf A}(t,{\bf k}) \cdot {\bf B}^*(t,{\bf k})>0$.
From Eq.~(\ref{eq:anomaly}),
this unstable mode decreases $\mu_5$ for fixed $T$ and 
increases the magnitude of ${\cal H}$. 
As a result, it weakens the instability.

The concrete time evolution of the plasma beyond the leading order in 
$A^{\mu}$ can be described by the full kinetic theory 
[see Eq.~(15) in Ref.~\cite{Son:2012zy}]
together with Eqs.~(\ref{eq:maxwell}), (\ref{eq:j}), and (\ref{eq:anomaly}), 
which would require a numerical analysis.
Still the saturation of the instability itself may be understood from 
the energy and helicity conservations. 
We assume $\mu_5 \sim T$ or $\mu_5 \gg T$.
As the energy and magnetic helicity of electromagnetic fields, 
$\sim (kA)^2$ and $\sim \alpha kA^2$,
come from those of chiral fermions, 
$O(\mu_5^4)$ and $O(\mu_5^3)$,
we get the typical scales of $k$ and $A$ relevant to the instability:
\beq
\label{eq:scale_sat}
k \sim \alpha \mu_5, \qquad A \sim \frac{\mu_5}{\alpha}, \qquad B \sim \mu_5^2.
\eeq
The electromagnetic fields cannot grow beyond this and will saturate 
in the end. At the saturation, the gauge fields become nonperturbatively large.
Thus, if $\mu_5$ for electrons is produced during the evolution of neutron 
stars \cite{Charbonneau:2009ax}, chiral plasma instabilities would provide 
a mechanism to generate the large magnetic helicity which plays an important 
role for the stability of the large magnetic field (see, e.g., Ref.~\cite{Spruit:2007bt}).

\emph{Effect of collisions.}---%
For $\mu_5 \sim T$, the mean free time for electric charge transfer 
(or large-angle scatterings) of the plasma constituents in QED, 
$\tau_{\rm large}\sim 1/(\alpha^2 T \ln \alpha^{-1})$
\cite{Baym:1990uj}, is shorter than the time scale of the plasma instability, 
$\tau_{\rm inst} \sim 1/(\alpha^2 \mu_5)$.
On the other hand, the mean free path, $l_{\rm large} \sim \tau_{\rm large}$, 
is larger than the typical wavelength of the plasma instability, 
$1/k_{\rm inst} \sim 1/(\alpha \mu_5)$, and one expects that the effect of 
collisions is irrelevant for it.

To see this explicitly, consider the effect of collisions in the relaxation time 
approximation: we add the term $-\delta n_{\bf p}/\tau_{\rm rel}$ 
in the right-hand side of Eq.~(\ref{eq:simple_kin}), where $\tau_{\rm rel}$ is the the 
relaxation time assumed to be of the same order as $\tau_{\rm large}$.
Then $\omega$ in the denominator of Eq.~(\ref{eq:P-even}) is replaced by 
$\omega + i/\tau_{\rm rel}$. Repeating a similar computation to above, one finds that
the modification to Eq.~(\ref{eq:unstable}) by the effect of collisions is subleading in 
$1/(k_{\rm inst} \tau_{\rm rel}) \ll 1$ and is negligible. 
(From this argument, the effect of collisions is clearly negligible for $\mu_5 \gg T$.)
As we shall see below, however, this is not the case in QCD.

\emph{Quark-gluon plasma with $\mu_5$.}---%
Suppose $\mu_5$ ($\sim T$) for quarks is initially generated in the QGP before the
thermalization, whose evolution is described by the kinetic theory.
(The case without $\mu_5$ but with the anisotropy of $n_{\bf p}$ 
was analyzed in Refs.~\cite{Romatschke:2003ms, Arnold:2003rq}.)
The above arguments are then extended to color electromagnetic fields, 
within the leading order in $A$, by replacing 
$\alpha \rightarrow \alpha_s=g^2/(4\pi)$ and 
$m_D^2 \rightarrow (N_f + 2N_c)g^2 T^2/6 + N_f g^2 \mu^2/(2\pi^2)$
with $N_f$ and $N_c$ the number of flavors and colors. 
The qualitative difference from the QED plasma is that the mean free time (path) 
for color charge transfer (or small-angle scatterings) is now 
$\tau_{\rm small} \sim l_{\rm small} \sim 1/(\alpha_s T \ln \alpha_s^{-1})$
because colored gluons, exchanged between quarks, can change the 
color charges of the scatterers 
\cite{Selikhov:1993ns, Heiselberg:1994px, Bodeker:1998hm, Arnold:1998cy}.
As $\tau_{\rm small} \ll \tau_{\rm inst}$ and $\l_{\rm small} \ll 1/k_{\rm inst}$
in this case, the collisions can affect the plasma instability. 
In the relaxation time approximation, 
$L(\omega + i/\tau_{\rm rel}, k) \simeq -i \tau_{\rm rel}[1 - (k \tau_{\rm rel})^2/3]$ 
with $\tau_{\rm rel} \sim \tau_{\rm small}$, 
one indeed finds that Eq.~(\ref{eq:unstable}) is modified to
\beq
\label{eq:unstable_mod}
\qquad \gamma(k) = \frac{3 N_f \alpha_s \mu_5 k}{\pi m_D^2 \tau_{\rm rel}}
\left(1-\frac{\pi k}{\alpha_s \mu_5}\right).
\eeq
The unstable modes in $0\leq k\leq \alpha_s \mu_5/\pi$ still exist but grow 
more rapidly with $\tau_{\rm QCD} \sim 1/(\alpha_s^2 \mu_5 \ln \alpha_s^{-1})$
\cite{CMW}.

\emph{Conclusion and outlook.}---%
We have shown that the relativistic electromagnetic plasma and QGP with 
finite $\mu_5$ are dynamically unstable and $\mu_5$ is damped by the 
exponential growth of (color) electromagnetic fields at the early stage.
As the gauge field grows, the nonlinearity of the gauge field becomes important \cite{magnetic}.
To understand the roles of the chiral plasma instability at the later stage 
quantitatively, more detailed studies of the dynamical evolution of plasmas 
based on the kinetic theory need to be worked out. It would also be an 
interesting question whether the instability persists even at strong coupling 
where the kinetic description breaks down.

There are also other directions in which one can extend or improve our analysis.
(i) It is straightforward to include the anisotropy of $n_{\bf p}^0$
analogous to Ref.~\cite{Romatschke:2003ms}. 
This allows for the study of the coexistence (or competition) with the Weibel instability.
(ii) The effect of the quark mass can be incorporated in the chirality evolution 
in Eq.~(\ref{eq:anomaly}), which makes $\mu_5$ smaller. 
These analyses would be important to understand whether the CME
remains an observable effect in the realistic situations.

We thank T.~D.~Cohen, S.~Mahmoodifar, M.~E.~Shaposhnikov,
and D.~T.~Son for useful conversations, and K.~Fukushima for
comments.
N.Y. is supported by JSPS Research Fellowships for Young Scientists. 
N.Y. thanks the hospitality of the Quark-Hadron Theory Group and 
Kobayashi-Maskawa Institute for the Origin of Particles and the Universe 
(KMI) at Nagoya University, where this work was initiated.


\begin{thebibliography}{99}
\bibitem{mu_5}
It should be remarked that $\mu_5$ is not associated with any conserved 
charge owing to the axial anomaly, 
so that it cannot be treated as the usual chemical potential. 
Still $\mu_5$ can be introduced as an external parameter of a system 
if its variation is slow enough, as will be verified {\it a posteriori} below  
(see also Ref.~\cite{Kharzeev:2007tn}).
\end{thebibliography}
\end{document}